\documentstyle[preprint,epsf]{jpsj}
\def\EQ{\begin{eqnarray}}
\def\EN{\end{eqnarray}}
\def\bv{{\bf v}}
\def\str{\sigma^{p}}

\def\bvk{{\bf v_{\bf k}}}
\def\bvt{{\bf v_{\bf k}^\bot}}
\def\hW{\hat W}
\def\hG{\hat G}
\def\hF{\hat H}
\def\hE{\hat E}
\def\hLa{{\hat L}^G_{\bf k}}
\def\hLb{{\hat L}^H_{\bf k}}
\def\hSa{{\Sigma}^G_{\bf k}}
\def\hSb{{\Sigma}^H_{\bf k}}

\def\bfo{{\bf f}}
\def\bfk{{\bf f_{\bf k}}}
\def\bfl{{\bf f^{||}_{\bf k}}}
\def\bft{{\bf f^{\bot}_{\bf k}}}
\def\bk{{\bf k}}
\def\bp{{\bf p}}
\def\br{{\bf r}}
\def\tomega{{\tilde \omega}}
\inst{%
Department of Physics, Kyoto University, Kyoto 606-8502\\
}
\def\runtitle{Dynamic Critical Phenomena of Polymer Solutions}
\def\runauthor{Akira {\sc Furukawa}}

\title
{
Dynamic Critical Phenomena of Polymer Solutions
}

\author
{Akira {\sc Furukawa}}

\inst
{
Department of Physics, Kyoto University, Kyoto 606-8502\\
}

\recdate
{
\today
}
\recdate
{
\today
}

\abst{
Recently, a systematic experiment measuring critical anomaly of viscosity of polymer 
solutions has been reported by H. Tanaka and his co-workers 
(Phys. Rev. E, {\bf 65}, (2002), 021802). 
According to their experiments,
the dynamic critical exponent of viscosity $y_c$ drastically decreases with
increasing the molecular weight.  
In this article the kinetic coefficients renormalized by 
 the non-linear hydrodynamic interaction 
are calculated by the mode coupling theory. 
We predict that the critical divergence
of viscosity should be suppressed with increasing the molecular weight.   
The diffusion constant and the dynamic structure factor are also
calculated. The present results explicitly show that
the critical dynamics of polymer solutions should be affected by 
an extra spatio-temporal scale intrinsic to polymer solutions, and   
are consistent with the experiment of Tanaka, {\it et al}. 
\\
}
\kword{
dynamic critical phenomena, model-H universality, viscoelasticity, 
dynamical asymmetry coupling  
}

\begin{document}
\sloppy
\maketitle

\section{Introduction}
Dynamic critical phenomena of classical fluids have been successfully
investigated by the mode coupling theory
\cite{Kawasaki1,Kawasaki2,Kawasaki3} and the dynamic 
renormalization group theory
\cite{Halperin-Hohenberg}. 
Because of the divergence of the correlation length $\xi$ of 
concentration fluctuations, the kinetic coefficients exhibit 
critical anomaly. 
According to the mode coupling theory and the 
dynamic renormalization group theory
the anomalous kinetic coefficient for concentration behaves as 
$\epsilon^{-\nu x_\lambda}$
\cite{Kawasaki3,Halperin-Hohenberg},
where $\nu$ is the critical exponent of $\xi$ nearly equal to 0.63 
and $\epsilon=(T-T_c)/T_c$. Both theories predict that $x_\lambda\cong 0.95$.
Kawasaki\cite{Kawasaki2} and Perl and Ferrell \cite{Perl-Ferrell} 
predicted that the shear viscosity $\eta$ exhibits a
logarithmic divergence:
\EQ
\eta=\bar\eta\biggl[1+\frac{8}{15\pi^2}\ln(\Lambda_0\xi)\biggr],\label{log-d}
\EN  
where $\bar\eta$ is the bare viscosity and 
$\Lambda_0$ is a microscopic cut off wave number. 
Later, Ohta\cite{Ohta1} pointed out that the viscosity anomaly 
is multiplicative and behaves as
\EQ
\eta\sim\epsilon^{-\nu x_{\eta}}=\epsilon^{-y_c}.\label{exp-d}
\EN
The dynamic exponent $y_c$ has a universal number about 0.04
\cite{Ohta-Kawasaki,Garistro-Kapal,Halperin-Hohenberg-Siggia,Siggia-Halperin-Hohenberg}. 
One of the important notions in the studies of the static and 
dynamic critical phenomena is the universality. 
This insists that phenomena with the same
set of critical exponents form a universality class. 
Dynamic critical phenomena of classical fluids belong to 
the model H universality 
in the Hohenberg-Halperin classification\cite{Halperin-Hohenberg}. 

It has been believed that the critical dynamics of complex fluids
such as polymer solutions also belongs to the model H universality.
Contrary to this conventional understanding, some experimental works 
have questioned that the dynamic critical phenomena of polymer 
solutions can be really categorized into the model H universality
\cite{Tanaka-Miura,Tanaka1,Berg-Gruner,Gruner-Hebib-Greer}. 
In fact the dynamic exponent $y_c$ obtained experimentally for polymer 
solutions is smaller than that in classical fluids
\cite{Tanaka-Miura,Tanaka1,Berg-Gruner,Gruner-Hebib-Greer}.    
Recently H. Tanaka and his co-workers\cite
{Tanaka-Nakanishi-Takubo} have reported  
experimental work of dynamic critical phenomena of polymer solutions.
Their accurate measurements for critical anomaly of viscosity show that
the dynamic exponent $y_c$ 
significantly decreases as the molecular weight $M_w$ 
increases. That is, in polymer solutions with large molecular 
weight the critical divergence of viscosity is suppressed.  
They also measured the dynamic structure factor 
by the light scattering experiment. 
It has been found that 
in the case of large molecular weight the dynamic structure factor cannot 
be expressed by the Kawasaki scaling function even in the vicinity of
the critical point. 
On the basis of such experimental facts, 
they concluded that the critical dynamics of 
polymer solutions exhibits a non-universal nature and cannot be 
classified into the model H universality in a practical sense
\cite{Tanaka-Nakanishi-Takubo}. 
They supposed that such a nontrivial behavior is due to the 
dynamic coupling between critical concentration fluctuations and 
another slow viscoelastic mode which is intrinsic to polymer solutions.
It should be noted that 
the viscoelastic effect and the break-down of the model H universality 
have been definitely confirmed in the studies for the phase separations of 
polymer solutions and dynamically asymmetric polymer blends\cite{Tanaka2}. 

Motivated by the experiments reported 
by Tanaka, {\it et al}\cite{Tanaka-Nakanishi-Takubo}, 
we shall investigate the dynamic critical phenomena 
of polymer solutions by the mode coupling theory. 
In the present analysis 
we use the two-fluid model
\cite{Brochard-deGennes,Brochard-deGennes2,Doi-Onuki} as the 
basic equations describing the dynamics of polymer solutions.
The two-fluid model was put forward by Brochard and de Gennes
\cite{Brochard-deGennes,Brochard-deGennes2}. 
Later, the current form of the two-fluid model was 
derived by Doi and Onuki\cite{Doi-Onuki}.
It has been successfully used to explain various phenomena 
such as shear induced concentration fluctuations
\cite{Doi-Onuki,Onuki1,Helfand-Frederickson,Milner},
non-exponential decay in dynamic scattering near equilibrium
\cite{Brochard-deGennes,Brochard-deGennes2,Doi-Onuki,Adam-Delsanti},
phase separation dynamics\cite{Tanaka2,Onuki-Taniguchi,Taniguchi-Onuki,Onuki-book}, and so on.
However, the critical anomaly of polymer solutions has not been studied 
by the two-fluid model.

Although the viscoelastic effect on the relaxation
of the concentration fluctuations have been studied by a number of people 
within the framework of the two-fluid model 
(see for example ref. 19), 
the past studies were mainly performed in the systems where
the mode coupling is not effective on the thermal relaxations, 
such as one phase region far from the critical point, 
very viscous polymer solutions and polymer blends, and so on.
Therefore, in these analyses 
the velocity fluctuations have not been explicitly taken into consideration.
Consequently, they neglected the effect of the mode coupling on the 
thermal decay rates\cite{comment}. 
However, near the critical point in the semi-dilute polymer solutions, 
the dynamics of the critical fluctuations is strongly influenced by   
the non-linear hydrodynamic interaction arising from 
the streaming type mode coupling.
As a result, the kinetic coefficients, observed near the critical point,
are renormalized by the non-linear hydrodynamic interaction.   
In the present analysis such a non-linear hydrodynamic interaction is
successfully taken into account by the mode coupling theory.
We then show that 
the viscoelasticity affects the critical dynamics of polymer solutions,
resulting in a suppression of the divergence of viscosity.
This result properly explains the experiments reported by Tanaka, {\it et al}
\cite{Tanaka-Nakanishi-Takubo}.

The organization of the present paper is as follows:
In \S2 we briefly review the two-fluid model for the polymer solutions.
In \S3 we analyze the viscoelastic effect on the 
ralaxation of the concentration fluctuations and
the velocity fluctuations within the linearized two-fluid model equations
and derive the bare (unperturbed) propagators 
which give a starting point of the 
mode coupling analysis performed in \S4. 
Here we shall evaluate the complex shear viscosity 
that is non-local both in space and time, 
which results from the dynamical asymmetry coupling 
between the viscoelastic stress and the velocity fluctuations.
In \S4 we investigate the critical anomaly, 
constructing self-consistent equations for the propagators 
by means of the mode coupling theory.
The kinetic coefficients renormalized 
by the non-linear hydrodynamic interaction 
which is important in the vicinity of the critical point are calculated.    
In the final section we will give some remarks about the present analysis. 
\section{Basic equations}
In this article we use the two-fluid model for the basic equations
to describe the dynamics of polymer solutions. 
Here we briefly survey the two-fluid model equations. 
The readers who want to know the details of the two-fluid model, 
see the original paper\cite{Doi-Onuki} or excellent reviews
\cite{Tanaka2,Onuki-book}.

Let ${\bv}_p(\br,t)$ and ${\bv}_s(\br,t)$ be the average velocities 
of polymer and solvent, respectively, and $\psi(\br,t)$ is 
the volume fraction of polymer at point $\br$ and time $t$. 
We assume that the solution is incompressible and that polymer 
and solvent have the same specific volume. 
Thus $\psi$ satisfies a conservation law with $\bv_p$:
\EQ
\frac{\partial}{\partial t}\psi =-\nabla\cdot(\psi{\bv}_p).\label{conservation}
\EN
The volume-averaged velocity $\bv=\psi\bv_p+(1-\psi)\bv_s$ obeys 
the following hydrodynamic equation 
\EQ
\rho\frac{\partial {\bv}}{\partial t}=
-\nabla p+\eta_0\nabla^2{\bv}
-\psi\nabla\frac{\delta F}{\delta \psi}+\nabla\cdot{\hat\sigma}^p.
\EN
where $\rho$ is the average mass density, $\eta_0$ is the solvent viscosity,
$F$ is the free energy functional of the system, 
and ${\hat\sigma}^p$ is the viscoelastic stress.
The hydrostatic pressure $p$ is determined by the 
incompressibility condition
\EQ
\nabla\cdot{\bv}=0.
\EN
For slow motions, the two-fluid model gives the convective velocity 
of polymer $\bv_{p}$ as
\EQ
{\bv}_{p}&=&{\bv}+\frac{(1-\psi)^2}{\zeta(\psi)}
(-\psi\nabla\frac{\delta F}{\delta \psi}+\nabla\cdot{\hat\sigma}^p),\label{p-velocity}
\EN
where $\zeta(\psi)$ is the friction coefficient between the two components, 
and is given by
\EQ
\zeta(\psi)=6\pi\eta_0\xi_b^{-2}.
\EN
Here $\xi_b$ is the blob length of order $b/\psi$, with $b$ being the monomer size. 
Eqs. (\ref{conservation}) and (\ref{p-velocity}) imply that 
the non-vanishing viscoelastic stress 
produces a diffusion. This effect is called the stress-diffusion coupling.
The viscoelastic stress due to the network deformations is given by
\EQ
{\hat \sigma}^p=G(\psi)\hW\cdot(\hW-{\hat\delta}).
\EN
This form is derived from the free energy functional introduced 
in eq.(10) shown below.  
Here $G(\psi)$ is the $\psi$-dependent shear modulus, $\hat \delta$ denotes the unit matrix
and $\hW$ is interpreted as a long lived strain variable. 
We assume that the equation of motion of $\hW$ is 
given by the following dynamic equation\cite{Taniguchi-Onuki,Onuki-book}, 
\EQ
\frac{\partial \hW}{\partial t}+({\bv_p}\cdot\nabla)\hW&=&
(\nabla{\bv}_p^{\dagger}\cdot\hW+\hW\cdot\nabla{\bv}_p)
-\frac{1}{\tau}(\hW-\hat\delta),
\EN
where $\tau(\psi)$ is 
the $\psi$-dependent relaxation time of shear stress.

\section{Linear Response}
In this section we examine the effect of the viscoelasticity 
on the relaxation of the concentration fluctuations 
and the velocity fluctuations,  
based on the linearized two-fluid model equations.
We then obtain the bare (unperturbed) 
propagators which give the starting point of 
the mode coupling analysis performed in the next section. 
Previously, Doi and Onuki\cite{Doi-Onuki} 
showed that the diffusion process is drastically influenced 
by the dynamical asymmetry coupling ($\alpha\ne 0$) 
between the concentration fluctuations 
and the longitudinal mode of the viscoelastic force 
(stress-diffusion coupling). 
Their analysis predicted an additional wave number dependence 
of the diffusion coefficient due to the viscoelasticity, 
which was confirmed by the recent experimental studies 
(see for example ref. 27).
However, in their analysis 
the very viscous systems without the velocity fluctuations were assumed, 
so the effect of the viscoelasticity 
on the relaxation of the velocity fluctuations (hydrodynamic relaxation) 
was not investigated. 
Here we take the transverse parts of the viscoelastic force and 
the velocity fluctuations into consideration.
We shall then show that the dynamical asymmetry coupling 
between these transverse modes has a strong influence on 
the hydrodynamic relaxation, 
resulting in a wave number dependent shear viscosity.

The relaxation dynamics of the thermal fluctuations 
in a one phase region far from the critical point
is still described by the linearized equations. 
Since the longitudinal modes do not couple to the transverse modes
within the linearized equations,
the diffusion process far from the critical point 
should not be influenced by the hydrodynamic relaxation.
However, the situation is completely different 
in the vicinity of the critical point,
since the non-linear hydrodynamic interaction, 
resulting from the streaming type mode coupling, becomes dominant.
In the next section this will be shown 
by means of the mode coupling theory.

Now we shall investigate 
the relaxation dynamics of concentration fluctuations 
and velocity fluctuations 
within the linearlized approximation of 
the set of equations represented in \S2.
We set $\tau_0=\tau(\psi_0)$ and $G_0=G(\psi_0)$ 
with $\psi_0$ being the average concentration. 
The free energy functional is assumed to be of 
the following Ginzburg-Landau type\cite{Onuki1,Onuki-book}  
\EQ
F\{\psi,\hW\}=\int d{\bf r}[\frac{1}{2}r_0\psi^2
+\frac{1}{2}c_0|\nabla\psi|^2+\frac{1}{4}G_0\delta W_{ij}^2 ],
\EN
where $\delta W_{ij}=W_{ij}-\delta_{ij}$.
The Fourier transform of an arbitary function $g(\br)$ is defined by
\EQ
g_\bk=\int d\br e^{-i\bk\cdot\br}g(\br).
\EN
In terms of the Fourier components, the linearized equations are given by
\EQ
\rho\frac{\partial\bvt}{\partial t}&=&-\eta_0 k^2\bvt+
i({\hat\delta}-{\hat\bk}{\hat\bk})\cdot\bk\cdot{\hat\sigma}^p_\bk, \label{l1}\\
\frac{\partial}{\partial t}\delta\psi_{\bk}&=&-\Gamma_{\bk}\delta\psi_{\bk}
+\alpha L\bk\cdot\bk\cdot{\hat \sigma}_\bk^p, \label{l2} \\
\frac{\partial}{\partial t}\str_{\bk,ij}&=&-\frac{1}{\tau_0}\str_{\bk,ij}
+iG_0(k_iv_{\bk,j}^{\bot}+k_jv_{\bk,i}^{\bot})\nonumber\\
&&+2G_0\alpha\Gamma_{\bk}{\hat k_i}{\hat k_j}\delta\psi_{\bk}
-G_0\alpha^2Lk^2({\hat k_i}{\hat k_l}\sigma^p_{\bk,lj}
+{\hat k_j}{\hat k_l}\sigma^p_{\bk,li})\label{l3},\\
\str_{\bk,ij}&=&G_0\delta W_{\bk,ij}\label{l4},
\EN
where $\bvt$ is the transverse part of $\bvk$, $\hat\bk=\bk/k$, and
\EQ
\alpha=\frac{1}{\psi_0}
\EN
is the dynamical asymmetry parameter of the polymer solutions
\cite{Doi-Onuki,Onuki-Taniguchi}.
The kinetic coefficient $L$ is given by
\EQ
L=\psi_0^2(1-\psi_0)^2/\zeta(\psi_0).
\EN
The decay rate in the absence of the viscoelastic coupling is given by
\EQ
\Gamma_{\bk}=L(r_0+c_0k^2)k^2. 
\EN
Here and in the following the Einstein summation convention has been adopted. The viscoelastic force $\bfo$ is defined as follows:
\EQ
\bfo\equiv\nabla\cdot{\hat\sigma}^p.
\EN
For the present purpose it is convenient to decompose 
the Fourier transform ${\bfo}_\bk$ of the viscoelastic force 
into the longitudinal and the transverse components,
\EQ
\bfk={\hat{\bk}}{\hat{\bk}}\cdot\bfk+({\hat{\delta}}-{\hat{\bk}}{\hat{\bk}})\cdot\bfk.
\EN
Let us set
\EQ
\bfl\equiv&{\hat{\bk}}{\hat{\bk}}\cdot\bfk,\\
\bft\equiv&({\hat{\delta}}-{\hat{\bk}}{\hat{\bk}})\cdot\bfk.
\EN
From the linearized equations (\ref{l1})-(\ref{l4}), we obtain the following two sets of equations of motion for the longitudinal modes and the transverse modes.

For the longitudinal modes, we obtain
\EQ
\frac{\partial}{\partial t}\delta\psi_{\bk}&=&-\Gamma_{\bk}\delta\psi_{\bk}-\alpha LZ_k+R^{\psi}_{\bk}(t),{\label{L1}}\\
\frac{\partial}{\partial t}Z_{\bk}&=&-\frac{1}{\tau_0}(1+2\xi_{ve}^2k^2)Z_{\bk}-2G_0\alpha\Gamma_{\bk}k^2\delta\psi_{\bk}+R^{Z}_{\bk}(t),{\label{L2}}
\EN
where $Z_{k}=i\bk\cdot\bfl$ and $\xi_{ve}$ is 
a viscoelastic length or a magic length defined by
\EQ
\xi_{ve}\equiv\sqrt{G_0\tau_0\alpha^2L}.
\EN
This length was first introduced 
by Brochard and de Gennes\cite{Brochard-deGennes,Brochard-deGennes2}.
The physical significance of the viscoelastic length has been discussed 
by many researchers (see for example refs. 16 and 26).

For the transverse modes, we obtain
\EQ
\frac{\partial\bvt}{\partial t}&=&-\frac{\eta_0 k^2}{\rho}{\bvt}+\frac{1}{\rho}\bft+{\bf R}^{\bv}_{\bk}(t),{\label{T1}} \\
\frac{\partial}{\partial t}\bft&=&-\frac{1}{\tau_0}(1+\xi_{ve}^2k^2)\bft-G_0k^2{\bvt}+{\bf R}_{\bk}^{\bfo}(t).{\label{T2}}				\EN
Here $R^{\psi}_{\bk}$, $R^{Z}_{\bk}$,
${\bf R}^{\bv}_{\bk}$ and ${\bf R}^{\bfo}_{\bk}$ 
are random forces. 

First we formally solve eqs. (\ref{L1}) and (\ref{L2}) as
\EQ
\delta\psi_{\bk}(t)&=&G^{\psi\psi}_{\bk}(t)\delta\psi_{\bk}(0)+G^{\psi Z}_{\bk}(t)Z_{\bk}(0)\nonumber\\
&&+\int_{0}^{t}ds G^{\psi\psi}_{\bk}(t-s)R^{\psi}_{\bk}(s)+\int_{0}^{t}ds G^{\psi Z}_{\bk}(t-s)R^{Z}_{\bk}(s),\\
Z_{\bk}(t)&=&G^{Z\psi}_{\bk}(t)\delta\psi_{\bk}(0)+G^{Z Z}_{\bk}(t)Z_{\bk}(0)\nonumber\\
&&+\int_{0}^{t}ds G^{Z\psi}_{\bk}(t-s)R^{\psi}_{\bk}(s)+\int_{0}^{t}ds G^{Z Z}_{\bk}(t-s)R^{Z}_{\bk}(s),
\EN
where the propagators $G(t)$'s are written as
\EQ
G^{\psi\psi}_{\bk}(t)&=&\frac{1}{\omega_{+}-\omega_{-}}
\Bigl\{\bigl[\omega_{+}-\frac{1}{\tau_0}(1+2\xi_{ve}^2k^2)\bigr]\exp(-\omega_+t)
+\bigl[-\omega_{-}+\frac{1}{\tau_0}(1+2\xi_{ve}^2k^2)\bigr]\exp(-\omega_-t)\Bigr\}, \label{G0pp}\nonumber\\
&&\\
G^{\psi Z}_{\bk}(t)&=&\frac{\alpha L}{\omega_{+}-\omega_{-}}
\Bigl[\exp(-\omega_+t)-\exp(-\omega_-t)\Bigr],\\
G^{Z\psi}_{\bk}(t)&=&\frac{2G_0\alpha\Gamma_{\bk}k^2}{\omega_{+}-\omega_{-}}
\Bigl[\exp(-\omega_+t)-\exp(-\omega_-t)\Bigr],\\
G^{ZZ}_{\bk}(t)&=&\frac{1}{\omega_{+}-\omega_{-}}
\Bigl[(\omega_{+}-\Gamma_{\bk})\exp(-\omega_+t)
+(-\omega_{-}+\Gamma_{\bk})\exp(-\omega_-t)\Bigr].
\EN
The relaxation rates, $\omega_{+}$ and $\omega_{-}$ are given by
\EQ
\omega_{\pm}=\frac{1}{2}
\Biggl\{\bigl[\Gamma_{\bk}+\frac{1}{\tau_0}(1+2\xi_{ve}^2k^2)\bigr]
\pm{\sqrt{\bigl[\Gamma_{\bk}+\frac{1}{\tau_0}(1+2\xi_{ve}^2k^2)\bigr]^2-4\frac{\Gamma_\bk}{\tau_0}}}\Biggr\}.
\EN
The propagator $G^{\psi\psi}_{\bk}(t)$ is the dynamic structure factor, 
and it is expressed by a superposition of two exponential functions.
This form of the dynamic structure factor was first proposed by 
Brochard and de Gennes\cite{Brochard-deGennes,Brochard-deGennes2} 
on the basis of a phenomenological argument.
Afterwards, Doi and Onuki\cite{Doi-Onuki} 
derived the same form of the structure factor by analysing 
their two-fluid model equations which is employed in the present analysis 
(see also ref. 24).

Similarly as above, we can formally solve eqs. (\ref{T1}) and (\ref{T2}) and obtain 
\EQ
{\bvt}(t)&=&H^{\bv\bv}_\bk(t){\bvt}(0)+H^{\bv\bfo}_\bk(t)\bft(0)\nonumber \\
&&+\int_{0}^{t}ds H^{\bv\bv}_\bk(t-s){\bf R}^{\bv}_{\bk}(s)+\int_{0}^{t}ds H^{\bv\bfo}_\bk(t-s){\bf R}^{\bfo}_{\bk}(s),\\
{\bft}(t)&=&H^{\bfo\bv}(t){\bvt}(0)+H^{\bfo\bfo}_\bk(t)\bft(0)\nonumber \\
&&+\int_{0}^{t}ds H^{\bfo\bv}_\bk(t-s){\bf R}^{\bv}_{\bk}(s)+\int_{0}^{t}ds H^{\bfo\bfo}_\bk(t-s){\bf R}^{\bfo}_{\bk}(s).
\EN
where the propagators $H(t)$'s are given by
\EQ
H^{\bv\bv}_{\bk}(t)&=&\frac{1}{\Omega_{+}-\Omega_{-}}
\Bigl\{\bigl[\Omega_{+}-\frac{1}{\tau_0}(1+\xi_{ve}^2k^2)\bigr]\exp(-\Omega_+t)
+\bigl[-\Omega_{-}+\frac{1}{\tau_0}(1+\xi_{ve}^2k^2)\bigr]\exp(-\Omega_-t))\Bigr\}, \nonumber\\
&& \label{propagator-vv}\\
H^{\bv\bfo}_{\bk}(t)&=&-\frac{\rho^{-1}}{\Omega_{+}-\Omega_{-}}
\bigl[\exp(-\Omega_+t)-\exp(-\Omega_-t)\bigr],\\
H^{\bfo\bv}_{\bk}(t)&=&\frac{G_0k^2}{\Omega_{+}-\Omega_{-}}
\bigl[\exp(-\Omega_+t)-\exp(-\Omega_-t)\bigr],\\
H^{\bfo\bfo}_{\bk}(t)&=&\frac{1}{\Omega_{+}-\Omega_{-}}
\Bigl[(\Omega_{+}-\frac{\eta_0k^2}{\rho})\exp(-\Omega_+t)
+(-\Omega_{-}+\frac{\eta_0k^2}{\rho})\exp(-\Omega_-t)\Bigr].
\EN
The relaxation rates $\Omega_{+}$ and $\Omega_{-}$ are given by
\EQ
\Omega_{\pm}=\frac{1}{2}
\Biggl\{\bigl[\frac{\eta_0k^2}{\rho}+\frac{1}{\tau_0}(1+\xi_{ve}^2k^2)\bigr]\pm
\sqrt{\bigl[\frac{\eta_0k^2}{\rho}-\frac{1}{\tau_0}(1+\xi_{ve}^2k^2)\bigr]^2-\frac{4G_0k^2}{\rho}}\Biggr\}.
\EN

The viscoelasticity is characterized by the complex shear modulus 
$G^{*}_\bk(\omega)=G'_\bk(\omega)+iG''_{\bk}(\omega)$, 
where $G'_\bk(\omega)$ and $G''_\bk(\omega)$ are 
called the storage modulus and the loss modulus, respectively\cite{Doi-Edwards}.
From the Eqs.(\ref{T1}) and (\ref{T2}), 
the total shear stress ${\hat\sigma}_\bk(\omega)$ is expressed as
\EQ
{\sigma}_{\bk,ij}(\omega)=\frac{G^{*}_\bk(\omega)}{i\omega}\bigl[i k_i v^{\bot}_{\bk,j}(\omega)+i k_j v^{\bot}_{\bk,i}(\omega)\bigr],
\EN
where we have introduced the Laplace transforms of $\bvt(t)$ and ${\hat\sigma}_\bk(t)$ 
as
\EQ
\bvt(\omega)&=&\int_0^\infty dt\bvt(t)e^{-i \omega t}, \\
{\hat\sigma}_\bk(\omega)&=&\int_0^\infty dt{\hat\sigma}_\bk(t)e^{-i \omega t}. 
\EN
In the present system $G'_\bk(\omega)$ and $G''_\bk(\omega)$ are given by
\EQ
G'_{\bk}(\omega)&=&\frac{G_0\tau_0^2\omega^2}{\omega^2\tau_0^2+(1+\xi_{ve}^2k^2)^2},\\
G''_{\bk}(\omega)&=&\omega\bigl[\eta_0+G_0\tau_0\frac{1+\xi_{ve}^2k^2}{\omega^2\tau_0^2+(1+\xi_{ve}^2k^2)^2}\bigr],
\EN
which are non-local both in space and time. 
The complex shear viscosity is given by
\EQ
\eta_\bk(\omega)=\frac{G^{*}_\bk(\omega)}{i\omega}.
\EN
The imaginary part of $G^{*}_\bk(\omega)$ gives the shear viscosity 
for long time scale motion $(\omega\cong 0)$ as 
\EQ
\eta_\bk(0)=\lim_{\omega \to 0}\frac{G''_\bk(\omega)}{\omega}=\eta_0+\eta_p\frac{1}{1+\xi_{ve}^2k^2}, \label{shear-viscosity-das}
\EN
where we have defined $\eta_p\equiv G_0\tau_0$. 
The asymptotic behavior of $\eta_\bk(0)$ is represented by
\EQ
\eta_\bk(0)&\cong&\eta_0+\eta_p~~~~(k\xi_{ve}\ll 1),\\
&\cong&\eta_0+\eta_p\frac{1}{\xi_{ve}^2k^2}~~~~(k\xi_{ve}\gg 1)\label{shear-2}.
\EN 
For polymer solutions the two-fluid model gives the viscoelastic length as\cite{Doi-Onuki,Onuki-book} 
\EQ
\xi_{ve}\cong (\eta_p/\eta_0)^{\frac{1}{2}}\xi_{b}.\label{ve-blob}
\EN
Thus, Eq.(\ref{shear-2}) is also expressed as
\EQ
\eta_\bk(0)&\cong&\eta_0\bigl(1+\frac{1}{\xi_b^2k^2}\bigr),~~~~(k\xi_{ve} \gg 1),
\EN
which explicitly shows that the hydrodynamic interaction becomes weak 
beyond the blob length $\xi_{b}$.
It is worth mentioning that this property is due to 
the dynamical asymmetry coupling ($\alpha\ne 0$) 
between the viscoelastic stress and the velocity fluctuations.
For polymer solutions, it is well known that the diffusion process is 
drastically influenced by the dynamical asymmetry coupling
between the longitudinal modes, namely the so-called stress-diffusion coupling\cite{Brochard-deGennes,Brochard-deGennes2,Doi-Onuki}. 
The present analysis shows that 
the dynamical asymmetry coupling between 
the transverse modes is also important for 
the hydrodynamic relaxation process.  

\section{Mode coupling approach}
In this section we discuss the critical anomaly 
of the kinetic coefficients. 
If we take the whole non-linearity into consideration, many terms 
appear even in the second order 
in the perturbation expansion in terms of 
fluctuations and the analytical approach becomes very complicated.
However, it turns out that there are only a few non-linear terms 
which crucially affect the critical dynamics.
From the equations of motion represented in \S2, we find that
the non-linear terms, associated with the viscoelasticity, are so much smaller 
than the linear terms which are represented in the linearlized equations 
(\ref{l1})-(\ref{l4}) that we may safely discard the non-linear terms arising 
from the viscoelasticity, 
and retain only two streaming type mode coupling terms which 
arise from the hydrodynamic interaction:
One is the convection term of $\bv$ in the equation for $\psi$ and
the other is the osmotic pressure gradient term in the hydrodynamic equation.
We consider these nonlinearity up to the second order of the perturbation 
expansion.   
As a result the equations of motion that we must analyze are 
given as follows:
\EQ
\rho\frac{\partial\bvt}{\partial t}&=&-\eta_0 k^2\bvt+
i({\hat\delta}-{\hat\bk}{\hat\bk})\cdot\bigl[\bk\cdot{\hat\sigma}^p_\bk-\int_{\bp}\bp(r_0+c_0p^2)\psi_{\bk-\bp}\psi_{\bp}\bigr], \\
\frac{\partial}{\partial t}\delta\psi_{\bk}&=&-\Gamma_{\bk}\delta\psi_{\bk}
+\alpha L\bk\cdot\bk\cdot{\hat\sigma}^p_\bk-i\bk\cdot\int_{\bp}\psi_{\bp}\bv^{\bot}_{\bk-\bp},\\
\frac{\partial}{\partial t}\str_{\bk,ij}&=&-\frac{1}{\tau_0}\str_{\bk,ij}
+iG_0(k_iv^\bot_{\bk,j}+k_jv^\bot_{\bk,i})\nonumber\\
&&+2G_0\alpha\Gamma_{\bk}{\hat k_i}{\hat k_j}\delta\psi_{\bk}
-G_0\alpha^2Lk^2({\hat k_i}{\hat k_l}\sigma^p_{\bk,lj}
+{\hat k_j}{\hat k_l}\sigma^p_{\bk,li}),\\
\str_{\bk,ij}&=&G_0\delta W_{\bk,ij}.
\EN
As shown in the following sub-sections, our analysis based on the above 
set of equations well explains the experimental results
\cite{Tanaka-Nakanishi-Takubo}.
\subsection{Self-consistent Equations for Propagators}
In this sub-section we construct the self-consistent equations 
for the propagators within the mode coupling theory\cite{Kawasaki1,Kawasaki2}.
By the simplest approximation we obtain the 
following set of self-consistent equations for the propagators:
\EQ
\frac{\partial}{\partial t}\hG_\bk(t)&=&-\hLa\cdot\hG_\bk(t)\nonumber\\
&&-\frac{k_BT}{\rho}\int_0^tds\int_{\bp}k^2[1-(\hat\bp\cdot\hat\bk)^2]\frac{\langle|\psi_{\bk-\bp}|^2\rangle}{\langle|\psi_{\bk}|^2\rangle}H_\bp^{\bv\bv}(s)G_{\bk-\bp}^{\psi\psi}(s)\hE\cdot\hG_{\bk}(t-s),\label{Self1}\\
\frac{\partial}{\partial t}\hF_\bk(t)&=&-\hLb\cdot\hF_\bk(t)\nonumber\\
&&-\frac{k_BT}{4\rho}\int_0^tds\int_{\bp}p^2[1-(\hat\bp\cdot\hat\bk)^2]\times\nonumber\\
&&\langle|\psi_{\bk-\bp}|^2\rangle\langle|\psi_{\bp}|^2
\rangle(\frac{1}{\langle|\psi_{\bp}|^2\rangle}-\frac{1}{\langle|\psi_{\bk-\bp}|^2\rangle})^2G_\bp^{\psi\psi}(s)G_{\bk-\bp}^{\psi\psi}(s)\hE\cdot\hF_{\bk}(t-s),\label{Self2}
\EN
where $\langle\cdots\rangle$ denotes the equilibrium average, 
\begin{equation}
\hLa=
\begin{array}{r@{}l}
 &\left(
\begin{array}{ccc}
 L_{\bk}^{\psi\psi}&L_{\bk}^{\psi Z} \\
L_{\bk}^{Z\psi} & L_{\bk}^{ZZ}
\end{array} \right)=
 \left(
\begin{array}{ccc}
\Gamma_{\bk} & \alpha L\\
2G_0\alpha\Gamma_{\bk}k^2 & \frac{1}{\tau_0}(1+2\xi_{ve}^2k^2) 
\end{array} \right),
\end{array}
\end{equation}
\begin{equation}
\hLb=
\begin{array}{r@{}l}
 &\left(
\begin{array}{ccc}
 L_{\bk}^{\bv\bv}&L_{\bk}^{\bv \bfo} \\
L_{\bk}^{\bfo\bv} & L_{\bk}^{\bfo\bfo}
\end{array} \right)=
 \left(
\begin{array}{ccc}
\frac{\eta_0k^2}{\rho}& -\frac{1}{\rho}\\
G_0k^2& \frac{1}{\tau_0}(1+\xi_{ve}^2k^2)
\end{array} \right),
\end{array}
\end{equation}
and the matrix $\hE$ is given by
\begin{equation}
\hE=
\begin{array}{r@{}l}
 &\left(
\begin{array}{ccc}
 1&0 \\
0& 0
\end{array} \right).
\end{array}
\end{equation}
Equations (\ref{Self1}) and (\ref{Self2}) correspond to the diagrammatic 
one-loop approximation, neglecting vertex corrections.
\subsection{Renormalized Diffusion Constant}
First we calculate the diffusion constant and the 
dynamic structure factor in the vicinity of the critical point. 
The equations for the propagators can be written as
\EQ
\bigl[i\omega\hat\delta+\hLa+\hSa(\omega)\hE\bigr]\cdot\hG_\bk(\omega)&=&\hat\delta,\\
\bigl[i\omega\hat\delta+\hLb+\hSb(\omega)\hE\bigr]\cdot\hF_\bk(\omega)&=&\hat\delta,
\EN
where
\EQ
\hG_\bk(\omega)&=&\int_0^\infty dt e^{-i\omega t}\hG_\bk(t),\\
\hF_\bk(\omega)&=&\int_0^\infty dt e^{-i\omega t}\hF_\bk(t).
\EN
Here the self-energies $\hSa(\omega)$ and $\hSb(\omega)$ 
are given by
\EQ
{\hSa}(\omega)
&=&\frac{k_BT}{\rho}\int_0^\infty ds\int_{\bp}k^2[1-(\hat\bp\cdot\hat\bk)^2]\frac{\langle|\psi_{\bk-\bp}|^2\rangle}{\langle|\psi_{\bk}|^2\rangle}
H_\bp^{\bv\bv}(s)G_{\bk-\bp}^{\psi\psi}(s)e^{-i\omega s},\label{self-E1}\\
{\hSb}(\omega)
&=&\frac{k_BT}{4\rho}\int_0^\infty ds\int_{\bp}p^2[1-(\hat\bp\cdot\hat\bk)^2]\times\nonumber\\
&&\langle|\psi_{\bk-\bp}|^2\rangle\langle|\psi_{\bp}|^2
\rangle(\frac{1}{\langle|\psi_{\bp}|^2\rangle}-\frac{1}{\langle|\psi_{\bk-\bp}|^2\rangle})^2G_\bp^{\psi\psi}(s)G_{\bk-\bp}^{\psi\psi}(s)e^{-i\omega s}.
\label{self-E2}
\EN
These equations are so complicated that it is hopeless to solve
them analytically in general situations. 
However, near the critical point we are allowed to investigate eqs. 
(\ref{self-E1}) and (\ref{self-E2}) by the following iterative manner.
  
First we evaluate $\hSa(\omega)$ assuming that 
the shear viscosity does not exhibit any critical anomaly. 
As we shall see in the following analysis, 
the decay of $G_{\bk}^{\psi\psi}(t)$ is governed by the slowest mode 
near the critical point.
Therefore, $H_\bp^{\bv\bv}(s)$ in (\ref{self-E1}) is given by (\ref{propagator-vv}), so that we may set $G_{\bk-\bp}^{\psi\psi}(s)\cong 1$.
Considering the decay of concentration fluctuation, $\omega$ in (\ref{self-E1}) is much smaller than $\Omega_+$ or $\Omega_-$, whence we use the Markov
approximation to set $\omega=0$ in (\ref{self-E1}).
That is, our present analysis can be regarded as the
zeroth order approximation for the additional frequency
dependencies of the decay rates. 
If we use the Ornstein-Zernike form for $\langle |\psi_{\bk}|^2\rangle\propto
(k^2+\xi^{-2})^{-1}$, 
the resultant equations can be integrated appropriately as follows:
\EQ
\Sigma_\bk^G(0)&\cong&\int_{\bp}\frac{k_BTk^2}{p^2}\bigl[1-(\hat\bk\cdot\hat\bp)^2\bigr]\frac{1+\xi_{ve}^2p^2}{\eta_0+\eta_p+\eta_0\xi_{ve}^2p^2}\frac{1+\xi^2k^2}{1+\xi^2(\bk-\bp)^2}\nonumber\\
&=&\int_{\bp}\frac{k_BTk^2}{(\eta_0+\eta_p)p^2}\bigl[1-(\hat\bk\cdot\hat\bp)^2\bigr]\frac{1+\xi^2k^2}{1+\xi^2(\bk-\bp)^2}\nonumber\\
&&+\int_{\bp}\frac{k_BTk^2}{(\eta_0+\eta_p)p^2}\bigl[1-(\hat\bk\cdot\hat\bp)^2\bigr]\frac{\eta_p\xi_{ve}^2p^2}{\eta_0+\eta_p+\eta_0\xi_{ve}^2p^2}\frac{1+\xi^2k^2}{1+\xi^2(\bk-\bp)^2}\nonumber\\
&=&\frac{k_BT}{6\pi(\eta_0+\eta_p)\xi^3}\biggl[K(x)+\frac{\eta_p}{\eta_0}F(x,A)\biggr],\label{self-energy-G}
\EN
where $x=\xi k$, $k_B$ is the Boltzmann constant, 
and the parameter $A$ is given by
\EQ
A=\frac{\xi}{\xi_{ve}}{\sqrt{1+\frac{\eta_p}{\eta_0}}}.
\EN
By using eq.(\ref{ve-blob}), in the case $\eta_p\gg\eta_0$
$A$ is practically the ratio of the correlation length $\xi$ to 
the blob length $\xi_b$
\EQ
A\cong\frac{\xi}{\xi_b}.\label{A}
\EN
In eq.(\ref{self-energy-G}) the first term is the Kawasaki
scaling function, given by
\EQ
K(x)=\frac{3}{4}\biggl[1+x^2+(x^3-x^{-1})\tan^{-1}x\biggr].
\EN
In the present case 
an additional term $F(x,A)$ arises from the viscoelasticity,
which is given by
\EQ
F(x,A)&=&\frac{3}{4}x^2(1+x^2)\Biggl\{\frac{1}{2x^2}\biggl(1+\frac{1+x^2-A^2}{A}\biggr)+\biggl[\frac{x^2-1}{2x^3}-\frac{(1+x^2)^2-A^4}{4A^2x^3}\biggr]\tan^{-1}x\nonumber\\
&&-\frac{1+x^2-A^2}{2Ax^2}\biggl[1+\frac{(1+x^2-A^2)^2}{4A^2x^2}\biggr]
\bigg|\frac{2Ax}{1+x^2-A^2}\biggr|\tan^{-1}\bigg|\frac{2Ax}{1+x^2-A^2}\biggr|\nonumber\\
&&+\frac{(1+x^2)^2-A^4}{4A^2x(1+x^2)}\biggl[1+\frac{(1+x^2-A^2)^2}{4A^2x^2}\biggr]\frac{1}{{\sqrt{s(s+1)}}}\tan^{-1}\frac{\sqrt{s(s+1)}}{s}x\Biggr\},
\EN
where 
\EQ
s=\frac{(1+x^2-A^2)^2}{4A^2(1+x^2)}.
\EN
Though the above expression is very complicated, its asymptotic form 
becomes simpler:
\EQ
\Sigma_\bk^G(0)&\cong&
\frac{k_BT}{6\pi\xi^3(\eta_0+\eta_p)}x^2\bigl(1+\frac{\eta_p}{\eta_0A}\bigr),~~~~(k\ll\frac{1}{\xi}),
\label{As1}\\
&\cong&\frac{k_BT}{16\xi^3(\eta_0+\eta_p)}x^3\Bigl(1+\frac{8\eta_p}{3\pi\eta_0}\frac{x}{A}\Bigr),~~~~(\frac{1}{\xi}\ll k\ll\frac{1}{\xi_{b}}),\label{As2}\\
&\cong&\frac{k_BT}{16\xi^3\eta_0}x^3,~~~~(k\gg\frac{1}{\xi_{b}}).\label{As3}
\EN
Here we have neglected the terms irrelevant to the following analysis.
We plot the reduced self energy and its asymptotic form for various values of $\eta_p/\eta_0$ and $A$ in Figs. 1(a)-(f). 
We note that the dynamic scaling is violated and the self energy 
cannot be expressed by the Kawasaki scaling function.  
This is evident from Fig. 1(a) and Fig. 1(d) where the viscoelasticity is found to be 
relatively strong. 
 
The dynamic structure factor is given by
\EQ
G^{\psi\psi}_{\bk}(t)&=&\frac{1}{\tomega_{+}-\tomega_{-}}
\biggl\{\bigl[\tomega_{+}-\frac{1}{\tau_0}(1+2\xi_{ve}^2k^2)\bigr]
\exp(-\tomega_+t)
+\bigl[-\tomega_{-}+\frac{1}{\tau_0}(1+2\xi_{ve}^2k^2)\bigr]
\exp(-\tomega_-t)\biggr\}, 
\nonumber\\
\EN
where the renormalized relaxation rates, 
$\tomega_+$ and $\tomega_-$ are obtained as
\EQ
\tomega_{\pm}&=&\frac{1}{2}
\Biggl\{\Gamma_{\bk}+\hSa(0)+\frac{1}{\tau_{0}}(1+2\xi_{ve}^2k^2)\nonumber\\
&&\pm{\sqrt{\Bigl[\Gamma_{\bk}+\hSa(0)+\frac{1}{\tau_{0}}(1+2\xi_{ve}^2k^2)\Bigr]^2-4\frac{1}{\tau_0}\Bigl[\Gamma_{\bk}+\hSa(0)(1+2\xi_{ve}^2k^2)\Bigr]}}\Biggr\}.\label{renormalized-relaxation}
\EN  
A comment will be made on the above results.
Although the functional form of the dynamic structure factor 
$G^{\psi\psi}_{\bk}(t)$ is the same as eq.(\ref{G0pp}) 
derived in the previous section,
the relaxation rates $\tomega_\pm$ of the critical concentration fluctuations 
obtained here are the ones renormalized by 
the non-linear hydrodynamic interaction. 
  
In the vicinity of the critical point $\hSa(0)$ dominates $\Gamma_{\bk}$,
so the characteristic frequencies become
\EQ
\tomega_{+}&\cong&\frac{1}{\tau_0}(1+2\xi_{ve}^2k^2),\\
\tomega_{-}&\cong&\hSa(0).
\EN
The resultant structure factor is approximated as
\EQ
G^{\psi\psi}_{\bk}(t)\cong\exp[-\hSa(0)t].
\EN
That is, the critical mode associated with the 
concentration fluctuations can be free from the coupling 
with the viscoelasticity very near the critical point. 
Then, using eqs.(\ref{ve-blob}), (\ref{A}) and (\ref{As1}), 
the long-wavelength expression of the diffusion constant is given by
\EQ
D&=&\frac{k_BT}{6\pi\xi(\eta_p+\eta_0)}\Bigl(1+\frac{\eta_p}{\eta_0 A}\Bigr)\nonumber\\
&\cong&\frac{k_BT}{6\pi\xi(\eta_p+\eta_0)}\Bigl[1+\Bigl(\frac{\eta_p}{\eta_0}\Bigr)\Bigl(\frac{\xi_b}{\xi}\Bigr)\Bigr]\nonumber\\
&\cong&\frac{k_BT}{6\pi\xi(\eta_p+\eta_0)}\Bigl[1+\Bigl(\frac{\eta_p}{\eta_0}\Bigr)^{\frac{1}{2}}\Bigl(\frac{\xi_{ve}}{\xi}\Bigr)\Bigr].\label{long-wavelength-diffusion}
\EN

We point out that 
the temperature regime where the hydrodynamic interaction dominates
the diffusion process ($\hSa(0)\gg \Gamma_\bk$) becomes narrower as 
the molecular weight increases, 
since $\eta_p\propto M_w^3$\cite{Doi-Edwards}. 
\subsection{Renormalized Viscosity}
In the previous sub-section, we assumed that the shear viscosity 
does not show the critical anomaly. 
However, as is well known, 
the shear viscosity exhibits a weak divergence at the critical point.
In this sub-section, we investigate the critical anomaly of the
shear viscosity.
We can evaluate the self energy $\hSb(0)$ by putting $\omega=0$ in
(\ref{self-E2}): 
\EQ
\hSb(0)&=&\frac{k_BT}{4\rho}\int_0^{\infty}ds\int_{\bp}p^2[1-(\hat\bp\cdot\hat\bk)^2]\times\nonumber\\
&&\langle|\psi_{\bk-\bp}|^2\rangle\langle|\psi_{\bp}|^2
\rangle(\frac{1}{\langle|\psi_{\bp}|^2\rangle}-\frac{1}{\langle|\psi_{\bk-\bp}|^2\rangle})^2G_\bp^{\psi\psi}(s)G_{\bk-\bp}^{\psi\psi}(s).
\label{self-Markov2}
\EN
Retaining the lowest power of $k$, the resultant integration can be performed analytically, using the results obtained in the previous sub-section:
\EQ
\frac{\rho}{k^2}\hSb(0)&\cong&\frac{k_BT\xi^4}{30\pi^2}\int_0^{\Lambda_0}dp\frac{p^6}{(1+\xi^2p^2)^2}\frac{1}{{\Sigma}^G_{\bp}(0)},
\EN
where $\Lambda_0$ is the microscopic cut-off wave number. 
The above integrand can be estimated as 
\EQ
\frac{k_BT\xi^4}{30\pi^2}\frac{p^6}{(1+\xi^2p^2)^2}\frac{1}{{\Sigma}^G_{\bp}(0)}&\cong&\frac{\eta_0+\eta_p}{5\pi(1+\frac{\eta_p}{\eta_0 A})}\xi^5p^4,~~~~(p\ll\frac{1}{\xi}),\\
&\cong&\frac{8(\eta_0+\eta_p)}{15\pi^2}\frac{1}{p+\tau p^2},~~~~(\frac{1}{\xi}\ll p\ll\frac{1}{\xi_{b}}),\\
&\cong&\frac{8\eta_0}{15\pi^2}\frac{1}{p},~~~~(p\gg\frac{1}{\xi_{b}}),
\EN
where $\tau=8\eta_p\xi/3\pi\eta_0A $. 
In the case $\eta_p\gg\eta_0$ eqs.(\ref{ve-blob}) and (\ref{A}) give 
\EQ
\tau\cong\Bigl(\frac{\eta_p}{\eta_0}\Bigr)^{\frac{1}{2}}\xi_{ve}\cong\Bigl(\frac{\eta_p}{\eta_0}\Bigr)\xi_{b}.
\EN
The anomalous part of the shear viscosity $\Delta\eta$ is given by 
\EQ
\Delta\eta=\frac{\rho}{k^2}\hSb(0).
\EN
Thus, we obtain the total shear viscosity 
in the following form,  
\EQ
\eta&=&\bar\eta+\Delta\eta\nonumber\\
&\cong&\bar\eta\Bigl\{1+\frac{8}{15\pi^2}\bigl[\frac{3\pi}{40}\frac{1}{1+\frac{\eta_p}{\eta_0 A}}+\ln(\frac{\xi+\tau}{\xi_b+\tau})+\frac{\eta_0}{\bar\eta}\ln(\xi_b\Lambda_0)\bigr]\Bigr\},\label{shear viscosity of critical polymer solutions}
\EN
where $\bar\eta=\eta_0+\eta_p$ is the bare shear viscosity. 
In Fig.2 we plot the reduced shear viscosity
$\Delta\eta/\bar\eta=\eta/\bar\eta-1$ as a function of  $\eta_p/\eta_0$.
It is evident that the critical divergence of the shear viscosity
is suppressed as the molecular weight increases.
 
For $\eta_p\gg\eta_0$ and $\xi_{ve}\cong\xi$, it is readily shown that 
the resultant renormalized shear viscosity does not exhibit critical anomaly.  
\EQ
\frac{\eta}{\bar\eta}\cong 1.
\EN

For $\eta_p\cong\eta_0$ and  $\xi\gg\xi_{ve}$, 
where the viscoelasticity is relatively weak, 
we obtain the anomalous shear viscosity as follows: 
\EQ
\frac{\eta}{\bar\eta}\cong 1+\frac{8}{15\pi^2}\ln\Biggl[\frac{\xi\Lambda_0}{(\xi_{b}\Lambda_0)^{\frac{\eta_p}{\bar\eta}}}\Biggr].\label{log-dp}
\EN
Because of the small coefficient of the logarithmic term, 
it may well be exponentiated with the small exponent 
$x_{\eta}=8/15\pi^2\cong0.054$ as
\EQ
\frac{\eta}{\bar\eta}\cong \Biggl[\frac{\xi\Lambda_0}{(\xi_{b}\Lambda_0)^{\frac{\eta_p}{\bar\eta}}}\Biggr]^{x_{\eta}}.\label{exp-dp}
\EN
In the low molecular weight limit ($\eta_p=0$)
the expressions reduce to those of 
the simple classical fluids, with $y_c=x_{\eta}\nu\cong0.034$\cite{comment2}. 

The above results are due to the existence of an extra length-scale $\xi_{ve}$ 
intrinsic to the viscoelasticity. 
As is well known\cite{Kawasaki1,Halperin-Hohenberg}, 
in the case of classical fluids 
the behavior of the critical anomaly of the shear viscosity 
does not depend on the detail of the material. 
That is, the behavior of the critical anomaly 
does not depend on the material parameter, and is universal (model H). 
Contrary to this, as shown in the present analysis, 
in the case of polymer solutions 
the critical divergence of the shear viscosity strongly depend on 
the molecular weight $M_{w}$. 
Now, a few comments must be made on those points.
When the correlation length $\xi$ is very large compared with the other
length scale, 
it is shown that our results certainly exhibit the universal behavior. 
However, as noted by Tanaka\cite{Tanaka-Nakanishi-Takubo}, 
it is difficult to experimentally access 
such a temperature region for polymer solutions whose polymer has 
a very high molecular weight. 
In this sense we can say that the model H universality does not hold, 
at least in a {\it practical sense}\cite{Tanaka-Nakanishi-Takubo}.

\section{Conclusions and Remarks}
In this article we have investigated
the dynamic critical phenomena of polymer solutions.
Based on the mode coupling theory, we derived a set of
self-consistent equations for the self energies, 
taking the non-linear hydrodynamic interaction into account.
Renormalizing the non-linear hydrodynamic interaction terms 
which are important in the vicinity of the critical point, 
the transport coefficients were calculated.

Our analysis predicts that the critical divergence of the shear viscosity
is suppressed as the molecular weight increases.
This effect was indeed observed experimentally
by Tanaka, {\it et al}\cite{Tanaka-Nakanishi-Takubo}.
At the present stage the quantitative comparison with the experiments is
difficult in the following reasons.
There are many unknown parameters (for instance, $\eta_p$ and $\xi_b$ etc.) 
which cannot be evaluated explicitly. 
In addition, in their paper\cite{Tanaka-Nakanishi-Takubo} 
they introduced an effective exponent $y_c$ 
and then discussed the singularity of the shear viscosity. 
However, this does not relate directly to our result, 
eq. (\ref{shear viscosity of critical polymer solutions}).    

We have also calculated the diffusion constant 
eq.(\ref{long-wavelength-diffusion})
and the dynamic structure factor
$G_\bk^{\psi\psi}(t)$ with the renormalized relaxation rates 
$\tomega_{\pm}$, eq.(\ref{renormalized-relaxation}), 
by the non-linear hydrodynamic interaction. 
Those are new results for critical polymer solutions. 
I expect that the present theoretical analysis 
would stimulate further experimental studies.

In the present analysis we have used the Markov approximation in order
to
evaluate the self-energies.
Previously several authors investigated the memory effects in the
critical dynamics of classical fluids and predicted that
the effect has small contributions to the kinetic coefficients
\cite{Ohta2,Ferrell-Bhattacharjee}.
These theoretical predictions have been successfully confirmed
by the experiments\cite{Burstyn-Chang-Sengers}.
In the case of polymer solutions the separation of the time scale
between concentration fluctuations and the velocity is not clear
as in the case of classical fluids. Hence there is a possibility that
the viscoelasticity amplifies the frequency dependence of the kinetic
coefficients. We will present more detailed
studies for this elsewhere\cite{Furukawa}.

Finally, we note the following.
It is well known that the dynamical asymmetry coupling ($\alpha\ne 0$) between
the concentration fluctuations and the viscoelastic stress gives
the so-called stress-diffusion coupling\cite{Brochard-deGennes,Doi-Onuki}.
In addition to this effect, 
by the present linear response analysis (\S3),
it has become apparent that the hydrodynamic relaxation process
is also drastically influenced by the dynamical asymmetry coupling
between the velocity field and the viscoelastic stress.
The calculated bare shear viscosity has a wave number dependence
as seen from eq.(\ref{shear-viscosity-das}).

\section*{Acknowledgement}
The author wishes to thank Prof. Akira Onuki for useful comment.

\pagebreak

\section*{Figure Captions}
Fig.1. 
Reduced self energy 
$Q(x,A,\eta_p/\eta_0)=[K(x)+\frac{\eta_p}{\eta_0}F(x,A)]/x^2$ from eq.(\ref{self-energy-G}) (solid line).
The asymptotic form (broken line) is also plotted. 
We note that the dynamic scaling is broken down and the self energy 
cannot be expressed by the Kawasaki scaling function (dotted line),  
that is definite in (a) and (d) where the viscoelasticity is
relatively strong.\\

Fig.2. 
Reduced shear viscosity $\Delta\eta/\bar\eta=\eta/\bar\eta-1$ vs $\eta_p/\eta_0$ for various values 
of $A$. Here we set $\Lambda_0\xi_b=10$ and use eq.(\ref{A}). 
It is evident that 
the divergence of the shear viscosity is suppressed as the molecular weight is increased.


\begin{thebibliography}{25}
\bibitem{comment2}
In the same way, 
in the study of the dynamic critical phenomena of classical fluids,  
exponentiating eq.(\ref{log-d})
can give the dynamic exponent of the shear viscosity\cite{Ohta1}, 
which was first introduced in the study 
by the dynamic renormalization group theory\cite{Halperin-Hohenberg}.
Because the second term is smaller than the first in Eq.(\ref{log-d}), 
such a procedure is allowed 
under actual experimental conditions ($\xi\stackrel{<}{\sim} 10^{-3}$cm). 
However, there has been no definite conclusion 
which of the logarithmic divergence or the
exponential divergence is correct. 
By the way, in the present study, we estimate the viscosity anomaly 
as eq.(\ref{shear viscosity of critical polymer solutions}).
Due to the viscoelasticity, 
the expression is more complex than 
that in the case of classical fluids, 
and the magnitude of each term strongly depends on parameters, 
such as $A$ and $\eta_p/\eta_0$.
Therefore, 
it is generally difficult to obtain the multiplicative form by exponentiating 
eq.(\ref{shear viscosity of critical polymer solutions}). 
Moreover, the meaning of the multiplicative anomaly of the shear viscosity 
is still unclear in the study of critical polymer solutions.
Hence, we leave eq.(\ref{shear viscosity of critical polymer solutions}) 
as it is.
\end{thebibliography}
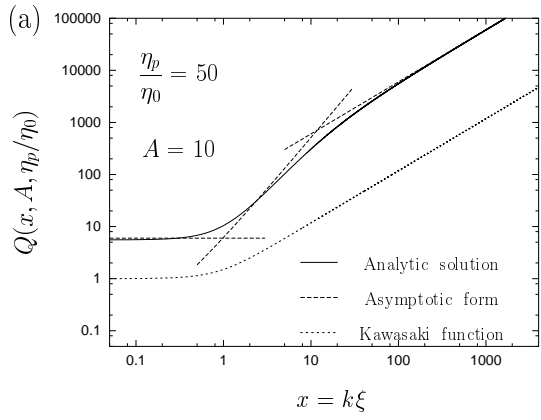
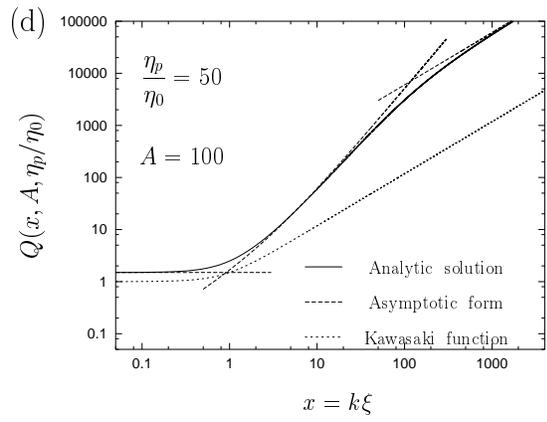
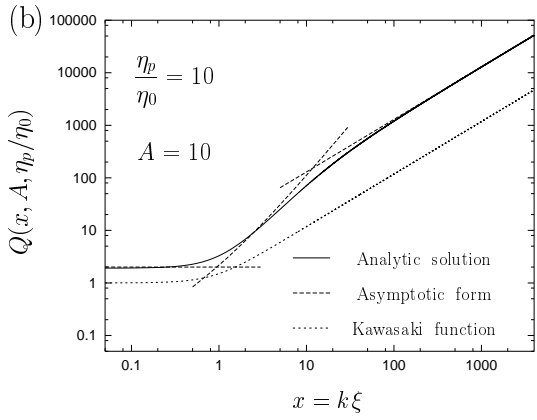
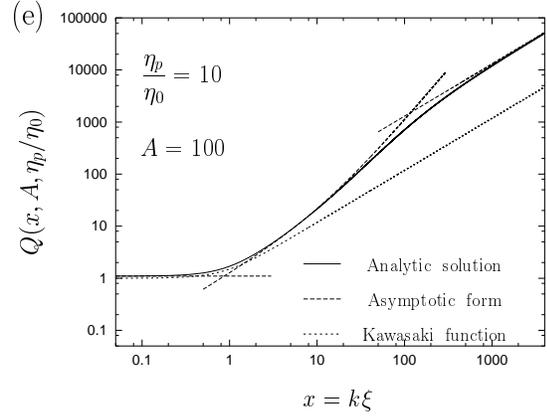
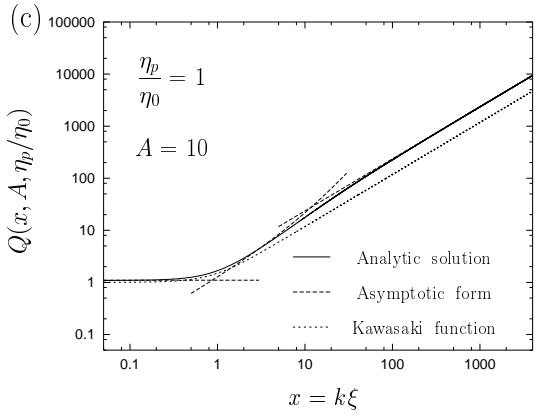
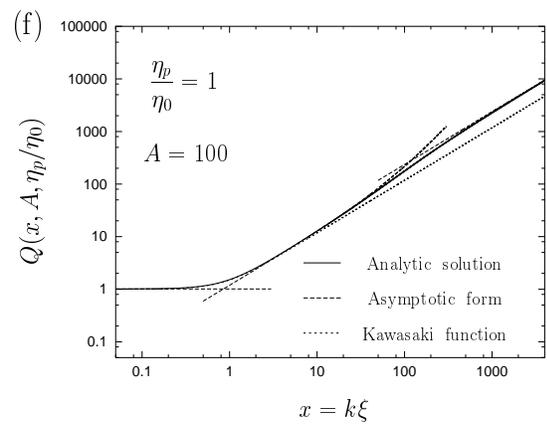
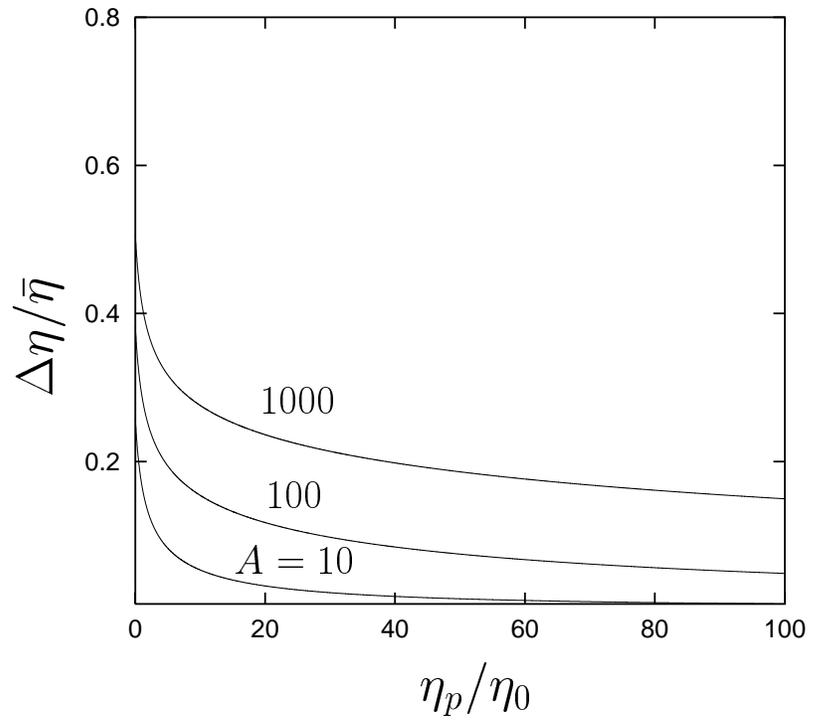
\end{document}